# Interfacial microscopic mechanism of free energy minimization in Ω precipitate formation

Sung Jin Kang[1,2,3], Young-Woon Kim[1], Miyoung Kim[1&], Jian-Min Zuo[2,3]*

[1] Department of Materials Science and Engineering and Research Institute of Advanced Materials, Seoul National University, Seoul 151-744, Korea.

[2] Department of Materials Science and Engineering, University of Illinois at Urbana-Champaign, Urbana, Illinois 61801, USA

[3] Frederick Seitz Materials Research Laboratory, University of Illinois at Urbana-Champaign, Urbana, Illinois 61801, USA.

*Corresponding author: jianzuo@illinois.edu

[&]E-mail: mkim@snu.ac.kr

**Abstract**

Precipitate strengthening of light metals underpins a large segment of industry. Yet, quantitative understanding of physics involved in precipitate formation is often lacking, especially, about interfacial contribution to the energetics of precipitate formation. Here, we report an intricate strain accommodation and free energy minimization mechanism in the formation of Ω precipitates ($Al_2Cu$) in the Al–Cu–Mg–Ag alloy. We show that the affinity between Ag and Mg at the interface provides the driving force for lowering the heat of formation, while substitution between Mg, Al and Cu of different atomic radii at interfacial atomic sites alters interfacial thickness and adjust precipitate misfit strain. The results here highlight the importance of interfacial structure in precipitate formation, and the potential of combining the power of atomic resolution imaging with first-principles theory for unraveling the mystery of physics at nanoscale interfaces.

Coherent precipitation in a crystalline matrix is accompanied by strain due to lattice misfits that fluctuate as discrete layers are added during growth. The misfit, which results from a difference in the lattice parameters between the precipitate and matrix, can give rise to large strain energy. If unmitigated, the strain energy, together with surface energies, drives phase changes in the commonly identified precipitation sequences, from coherent meta-stable phases to the incoherent thermodynamically stable phase accompanied by all-important alterations in mechanical properties[1,2]. The misfit, on the order of a fraction of the matrix d-spacing, includes a contribution from interface. Interface in general plays a critical role in the nucleation and growth of precipitates. Interfacial atomic structure and related physical properties can differ dramatically from neighboring bulk phases. However, determining or predicting interfacial atomic structure by experiment or theory alone is extremely difficult to do. For theory, the lack of precise atomic-scale composition information at or near the interface greatly increases the complexity of structure prediction problem, which limits its predictability. Without the structural knowledge, important questions remain unanswered about interfacial stability, interface processes during phase transformation, the role of interfacial intermixing and their contribution to coarsening resistance and interfacial strain.

Here, we report on a study of atomic structure and energetics of the interface formed between the $\Omega$ phase ($Al_2Cu$) and Al matrix in the Al–Cu–Mg–Ag alloy. Using a combination of first principles theory and 3D atomic resolution imaging, we determine the structure of a double-layer interface that forms at the broad face of $\Omega$ precipitates. We then calculate the lattice misfit and the heat of formation as function of interfacial composition. A free energy model based on these data then allows a comparison between the calculated and measured precipitate misifts and provides an explanation of the presence of Cu observed at the interface based on the interfacial free energy minimization mechanism.

The Al–Cu–Mg–Ag alloy belongs to the Al alloy 2000 series. Minor addition of Ag (~0.1 at.%) into the Al-Cu-Mg system with a high Cu/Mg ratio (10 to 1) drastically enhances the formation of uniformly-dispersed $\Omega$ phase with a large width-thickness aspect ratio (~$10^2$) on the {111} planes of the Al matrix (FCC, $\alpha$ phase) [3-6]. The presence of the $\Omega$ phase leads to a higher thermal stability, increased yield strength, and coarsening resistance, which has attracted considerable interests for aerospace and defense applications[2, 7, 8]. The $\Omega$ phase has an orthorhombic unit cell with $a_\Omega$=4.96 Å, $b_\Omega$=8.56 Å, $c_\Omega$=8.48 Å[6]. It can be regarded as a distorted form of the structure of tetragonal $\theta$-Al$_2$Cu precipitates found in over-aged Al-Cu alloys according to Knowles and Stobbs[6]. The $a_\Omega$ and $b_\Omega$ are 3 times the d-spacing of (211) and (110) of Al ($a_\alpha$=4.05 Å), respectively, with mismatch less than 0.015%. Along the c-axis, normal to the Al habit plane, the mismatch is large (-9.3% for one $\Omega$ unit cell) and it varies widely with thickness, contributing to a tensile or compressive strain in the Al matrix depending on the misfit. Composition analysis revealed strong segregation of Ag and Mg at the interface[9, 10]. The segregation has been suggested to be driven by the misfit strain energy because Ag and Mg atoms at the $\alpha/\Omega$ interface modifies the lattice, which could satisfy the minimum misfit condition at the Al{111} planes[11]. Several groups have attempted to determine the interfacial structure and its contribution to misfit[12-14]. However, the results are inconclusive. The resolution of previous experimental studies was not sufficient to resolve the atomic structure of the interface.

To gain insights about the interface, we have made new observations by examining the precipitates along four different orientations based on the epitaxial relationships of [100]Ω ∥[2-1-1]Al, [110]Ω∥[11-2]Al, [010]Ω∥[01-1]Al, and [310]Ω∥[1-10]Al. The image resolution is significantly improved by Cs correction[15, 16], which has only recently been applied to interface study in metals[17-19]. The interfacial atomic structure, as well as misfit, is determined based on the atomic resolution data and using *ab initio* density functional theory (DFT) based structure relaxation. Furthermore, DFT calculations of heat of formation and misfit are carried out. Together, they contribute to a free energy model based on substitution of Mg, Al and Cu at selected interfacial sites.

For the experiment, the aluminum alloy with the composition of Al–4%Cu–0.3%Mg–0.4%Ag by weight percentage was prepared in conditions according to the previous study [13]. Thin specimens were prepared by mechanical polishing, followed by ion-milling. Atomic resolution images in the so-called Z-contrast[20] were recorded using a JEOL 2200FS scanning transmission electron microscope (STEM) equipped with a CEOS probe aberration corrector. For composition analysis, electron energy loss spectroscopy (EELS) spectra were obtained using an in-column omega-energy filter for the energy dispersion, which is used for interfacial atomic identification.

Examples of the spatial averaged precipitate Z-contrast images at different sample orientations are shown in Fig. 1a-d. The averaging performed using the technique called template matching[21] improves the contrast of light atoms (Al especially). In all cases, the image resolution was sufficient to separate the projected atomic columns. For example, in the [310]Ω orientation, the Cu atomic columns are separated by 1.24 Å inside the precipitate according to the structural model [6]. The insets in Fig. 1a-d show the Z-contrast image simulation results based on the precipitate structure model we propose below. At the interface, strong contrast is seen in the recorded Z-contrast images with an atomic arrangement that is very different from that of $Al_2Cu$. The strong Z contrast is attributed to Ag, which has the highest atomic number (Z) in this alloy. Elemental analyses of the precipitate using EELS and energy dispersive X-ray spectroscopy (EDX) also revealed a single layer of Ag. In addition, STEM-EELS results indicate Cu atoms next to the Ag layer at the interface. Interfacial Mg is detected by EDX mapping (Details on composition analysis are provided in the supplemental materials).

Based on the above observations, we model the interface structure by first-principles DFT using the Vienna *ab-initio* simulation package (VASP) code[22, 23] and electron image simulations. The starting structure model, constructed based on the experimental data, has the interface made up of a layer of hexagonally arranged Ag in a graphene-like structure. In addition, three interfacial atomic sites next to Ag are included; they are labeled as A(I), A(II) and A(III) in Fig. 2. The A(I) site sits below the center of the Ag hexagon on the precipitate side. A(II) and A(III) form an separate atomic layer that resembles the structure of the Ω precipitate. Together, they make up the double-layer seen in the experimental Z-contrast images (Fig. 1). The proposed structure is relaxed using the conjugated gradient and residual minimization scheme–direct inversion in the iterative subspace (RMM–DIIS) algorithms [24]. The volume, atomic position, and unit cell shape are allowed to relax freely. Figure 2 shows a model precipitate after the structure relaxation. The model includes two unit-cells of $Al_2Cu$ [6], the interface, and four layers of Al above and below the precipitate. During structure relaxation, the A(I) site occupied by Cu or Mg atom, initially placed in middle of the Ag hexagon, moved down slightly towards the precipitate. This feature successfully reproduces features in the experimental images. Immediately below, on the A(II) atomic site, Cu atom placed on this site moved so that after structure relaxation it is positioned almost in the same plane as the A(III) atom site. Z-contrast image simulations using this relaxed structure with Cu occupying all three sites are shown in Fig. 1. The match between the relaxed theoretical structure and the experimental images at the interface are especially remarkable.

With the interfacial structure identified, we now can provide a description of the Ω precipitation sequence as it thickens. Figure 3 shows the atomic resolution Z-contrast images of three precipitates of 0, 0.5 and 2 unit cells thick Ω phase. In all cases, a double-layer interfacial atomic structure is seen on both sides of the precipitates. The thinnest precipitate is consisted of two double layers with an Al layer in the middle.

The interfacial composition is not uniform. As recorded atomic resolution Z-contrast images, as the example in Fig. 4a, show bright contrast at some of the A(I) sites, while other A(I) sites in darker contrast are likely occupied by light atoms of Al or Mg (The Z contrast difference between Al and Mg is too small to differentiate these two). Experimentally, it has been observed that the addition of Ag greatly enhances the formation of the Ω precipitates in the Al-Cu-Mg alloys, but Ag alone in the absence of Mg has no effect on the precipitation. EDX mapping using Mg $K_{1,2}$ indicates the presence of Mg atoms at the interface, in agreement with previous reports[9, 13]. The elevated Mg concentration at the interface was also reported in the 3D atom probe (3DAP) tomographic study [10, 25].

To investigate the effect of interfacial atomic substitutions, first we construct atomic models with the A(I) and A(II) sites substituted by Al or Mg. The model with the A(I) and A(II) sites occupied by Cu serves as a reference. Fig. 4b shows the effect of light atoms substitution. Substitution of Cu by Mg at A(I) leads to the largest reduction in the formation energy, while Mg at the A(II) site leads to an increase in the formation energy. The energy reduction/increase amounts to ~0.96 eV/atom for Mg at A(I) and 0.34~0.43eV/atom for Al at A(I) and A(II) respectively. Since the formation energy reduction by Mg at A(I) site is especially large, we expect significant Mg accumulation on the A(I) sites inside the interfacial layer. This result is fully consistent with the strong tendency toward Ag-Mg clustering observed in the 3DAP tomographic data [10, 25].

The substitution of light atoms also changes the precipitate thickness since the atomic radius of Mg (0.1602 nm) or Al (0.1432 nm) is larger than Cu (0.1278 nm). Figure 3b plots the calculated thickness of interface, Al and Al$_2$Cu layers (See Figure 2 for their definition). The changes in thickness of Al and Al$_2$Cu are small in these calculations. Large changes are observed in the thickness of interface. With Mg 100% at A(I), the modeled interface increases by 0.26 Å in thickness, compared with 100% Cu at A(I). Similarly, with 100% Al at A(II), the thickness increases by 0.17 Å.

The above results suggest the competing roles of formation energy and lattice misfit in the formation of Ω precipitates. To examine this, we construct a thermodynamic model for the precipitate free energy using the strain energy formulation of Lee, Barnett and Aaronson for a coherent ellipsoidal precipitate[26], which gives:

$$\Delta G / A = 2\mu_\alpha \varepsilon(c)^2 \frac{1+\nu_\alpha}{1-\nu_\alpha} f(\beta, \mu_\Omega / \mu_\alpha) t + \Delta H(c) - T\Delta S(c) \qquad (1)$$

Where $\Delta G$ is the change in free energy with Mg or Al substitution at A(I) or A(II) sites, A for the surface area and t for the precipitate thickness. The first term is the strain energy, where $c$ stands for interfacial composition, $\varepsilon$ the misfit, $\mu_\alpha$ and $\nu_\alpha$ are the shear modulus and Poisson ratio of the Al matrix, and $t$ the precipitate thickness. The value of the factor $f$ depends on the aspect ratio ($\beta$) and the shear modulus ratio $(\mu_\Omega / \mu_\alpha)$ and anisotropic elastic constants of the precipitate, which is largely unknown. Consequently, $f$ will be treated as a parameter here. For simplicity, we will consider substitution at A(I) site by Mg alone in what follows. There are ~11 A(I) sites on a 1 nm$^2$ precipitate interfacial area and the largest TΔS contribution at 473.15K is ~ 0.3eV/nm$^2$, which is small and will be neglected here. The $\varepsilon(c)$ and $\Delta H(c)$ are linearized using the data in Fig. 4.

Figure 5a plots the change in free energy per nm$^2$ as function of Mg occupation on A(I) for a precipitate with 2 unit cells of Al$_2$Cu and $f = 1$ to $6$. The precipitate has a calculated misfit of -0.19% with 100% Cu at A(I), while 100% Mg at A(I) gives 1.25% misfit. Mg atom at A(I) gives rise to $\Delta H$ = -0.96eV/atom or $\Delta H$ =-10.56eV/nm$^2$. The thermodynamic model predicts a minimum in the free energy at intermediate Mg substitution when the strain energy contribution is large with $f \approx 2$ or above. The minimum shifts to lower Mg content as $f$ increases.

Figure 5b plots the obtained misfit from the measured precipitate thicknesses. We have measured the average precipitate thickness from the atomic resolution images using the distances between Al (111) planes above and below the precipitate. The Al (111) d-spacing recorded in the experimental image, which is subtracted off from the measured distances, was measured away from the precipitate and used for calibration. For the precipitate with 2 unit cells of Al$_2$Cu, the measured $\varepsilon(c)$=0.35% ±0.29%, which is close to the minimum in the free energy obtained when $f \approx 3$. This, together with the observation of substantial amount of Cu at the A(I) sites, strongly support that interfacial substitution operates as the strain accommodation and free energy minimization mechanism.

For the precipitate with 0 unit cells of Al$_2$Cu, the calculated misfit is substantially larger and negative for 100% Cu at the A(I) site. Substitution of Mg reduces the negative misfit. Further substitution of Al at the A(II) site will bring the calculated misfit close to the experimental value. Together with Al at A(II) and Mg at A(I), the interfacial thickness increases by ~0.4 Å. This amounts to a change 2.6% and 1.6% in misfit for the precipitate with 0.5 and 1.5 unit cells of $Al_2Cu$ respectively. The measured strain at ~+/-5% for these two types of precipitate is thus beyond the interfacial strain accommodation range. Consequently, a large free energy barrier is expected, which in addition to the barrier to nucleate a half unit cell step at the interface contributes to the excellent coarsening resistance seen at elevated temperatures about 200 °C.

It should be pointed out that Ag atoms arranged in the unusual hexagonal network at the interface of Ω phase we identified appears to be an important feature in some of other Al alloy precipitates as well. Recent reports on Al-Cu-Li and Al-Mg-Zn-Cu alloys [19, 27] suggested hexagonal atomic arrangement at the interface with Cu and Zn atoms instead of Ag atoms. In the middle of these Cu and Zn hexagons, Li and Mg atoms are observed. Thus, the elucidation of the relationship between interfacial atomic structure, formation energy and strain potentially can provide a pathway to the identification of a range of highly stable precipitate structures by DFT calculations.

In conclusion, our combined experimental and theoretical study shows that a highly stable double-layer interfacial structure separates the Ω phase from Al in the Al–Cu–Mg–Ag alloy. The outermost layer next to Al is composed of Ag atoms in graphene-like hexagonal sites and Mg or Cu atoms below the center of the hexagon. The interfacial Mg atoms greatly stabilize the interface structure and consequently the Ω phase on the Al {111} habit as well. Furthermore, atomic substitutions of light atoms, Al and Mg, at the interface mediate the misfit strain and minimize free energy. Overall, atomic resolution microscopy played a critical role in the elucidation of interfacial atomic structure, while theory enables a construction of thermodynamic model based on calculated formation energy, precipitate strain and dependence on atomic substitutions.

**Acknowledgments:** Atomic resolution electron microscopy and data analysis were carried out by SJK and JMZ at University of Illinois, Urbana-Champaign under the support of DOE BES DEFG02-01ER45923. Synthesis and computational work were carried out at SNU by SJK, SWK and MK supported by a grant from the National Research Foundation of Korea, funded by the Ministry of Education, Science and Technology (NRF 20120005637 and NRF 20120006644). JMZ is additionally supported by NSF DMR 1006077.

**Figure Captions:**

**Figure 1** Experimental and simulated atomic-resolution HAADF images viewed from the [100]Ω(a),[110](b),[010](c),[310](d) of Ω precipitate. The experimental images are periodically averaged parallel to the precipitate. The averaging improves the visibility of light Al atoms. Simulated images based the structural model proposed here are displayed inside the white box for the comparison.

**Figure 2** A 3D rendering of the atomic structure of Ω precipitate inferred from Figure 1 and after structural relaxation. The atoms are colored in white (Ag), red (Cu), blue (Al). Interfacial sites, A(I), A(II) and A(III) are labeled and colored separately.

**Figure 3** Selected HAADF images showing the precipitates of different thicknesses with their atomic structures illustrated on the right. The thinnest precipitate observed has 0 unit cell of Ω phase between the interfaces. The precipitates retain its characteristic interface structure in all cases.

**Figure 4** (a) A Z-contrast image of a Ω precipitate viewed long [100]Ω. The boxed section is magnified and shown in (b). The magnified image show strong Z contrast on some of A(I) sites that are marked by arrows. Dark contrast are seen at other A(I) sites. (c) Calculated layer thickness(Å) and formation energy(eV) as a function of Mg and Al substitution at A(I) and A(II) sites.

**Figure 5** a) A comparison between the measured and calculated lattice misfits. The measurements are for precipitates with 0 to 2 unit cells of Ω phase. The calculated misfits are for the 0 and 2 unit cells cases with A(I)/A(II) sites occupied by Cu/Cu, Mg/Cu and Mg/Al respectively. b) Calculated free energy curves as function of Mg on A(I) sites for selected f values for the 2 unit cell Ω precipitate. The box marks the measured misfit.

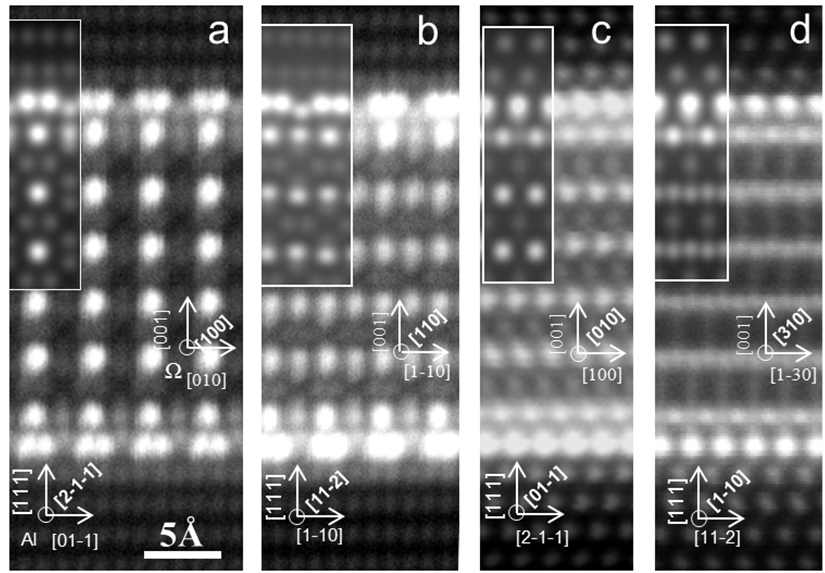

Figure 1

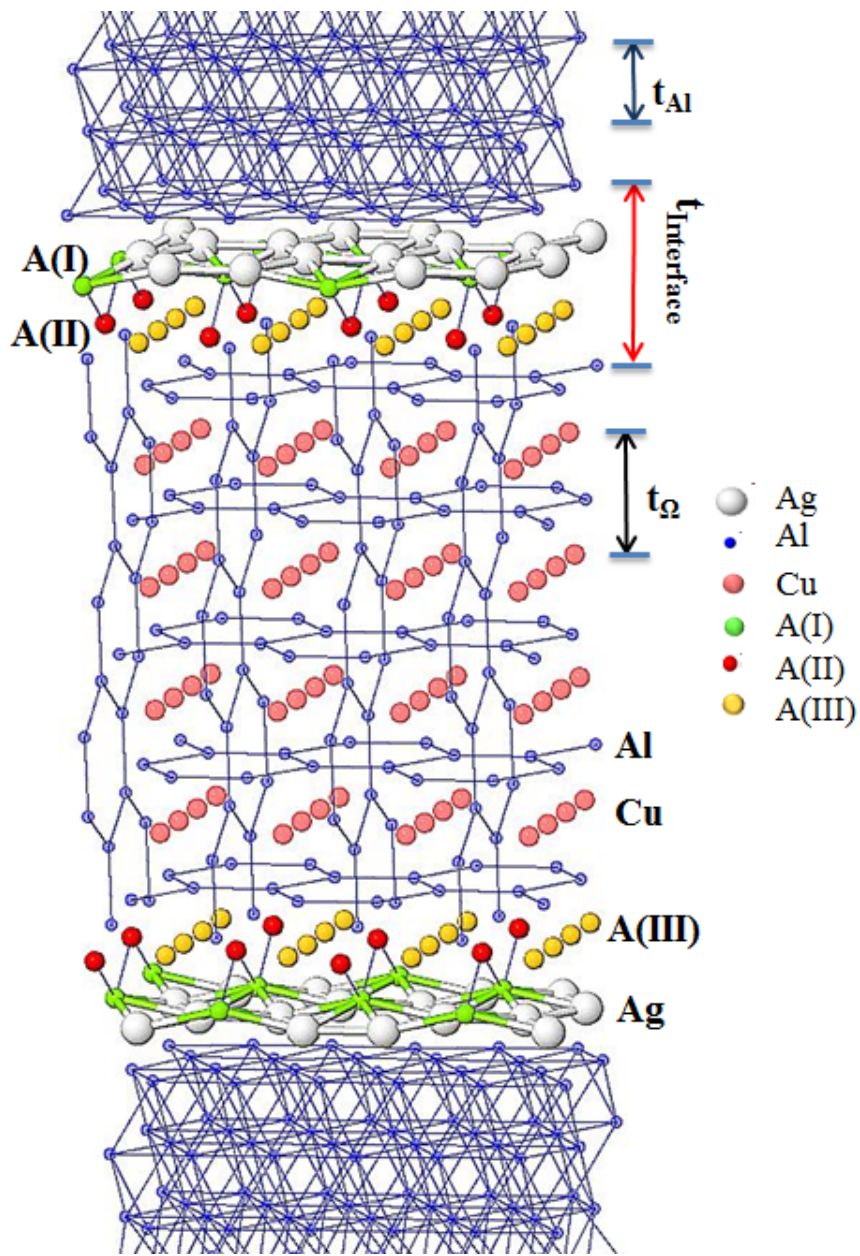

Figure 2

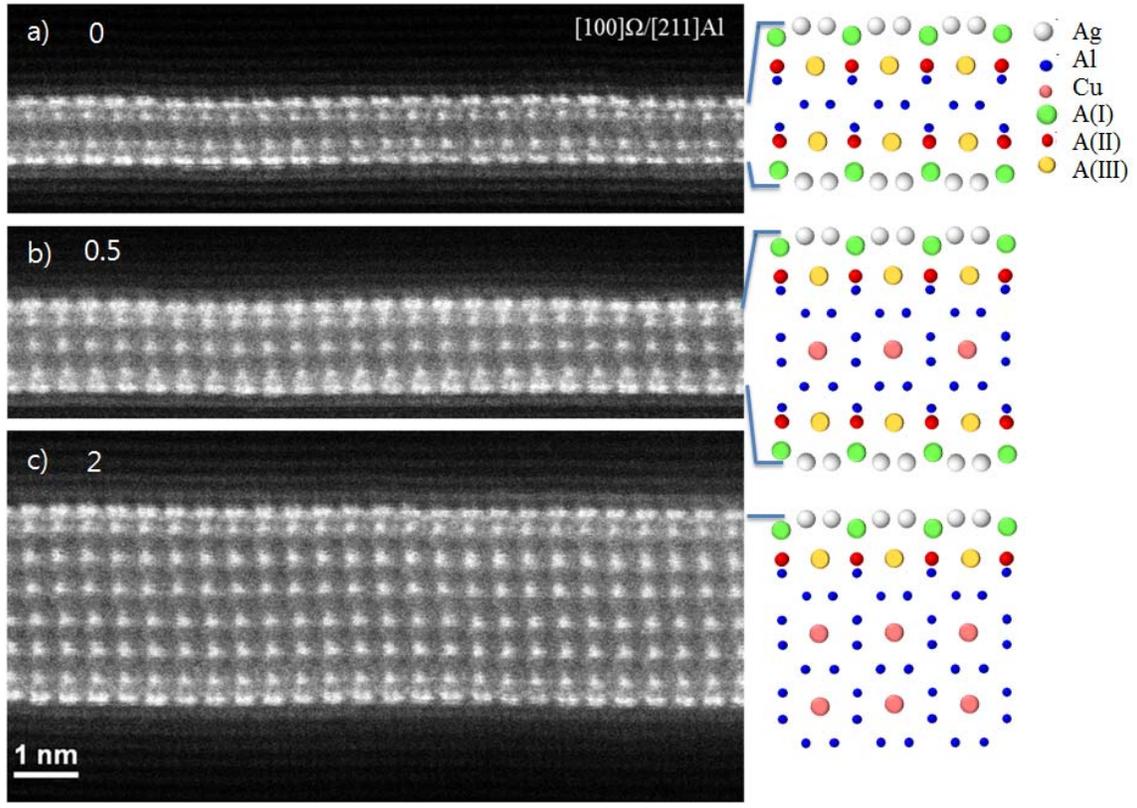

Figure 3

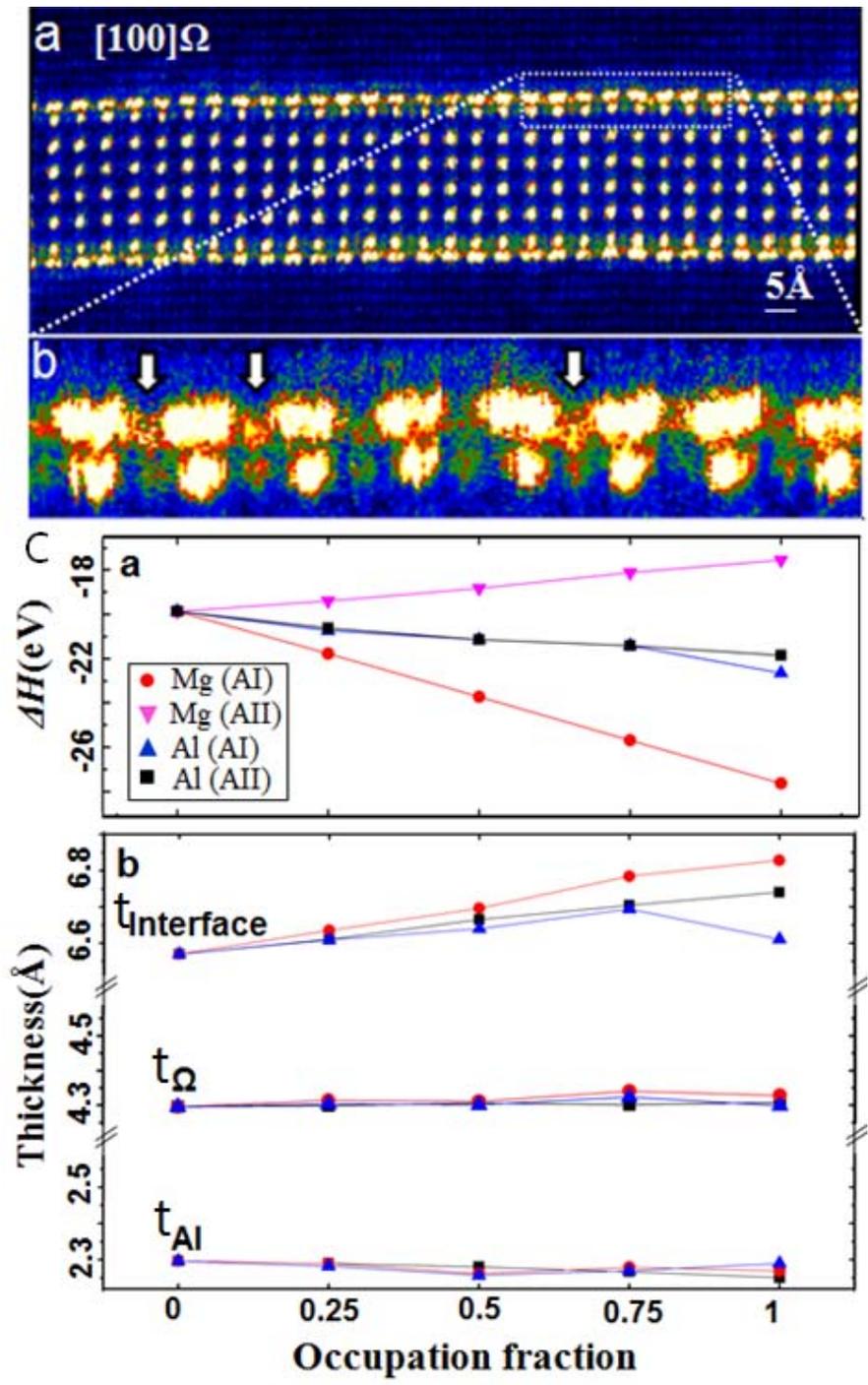

Figure 4

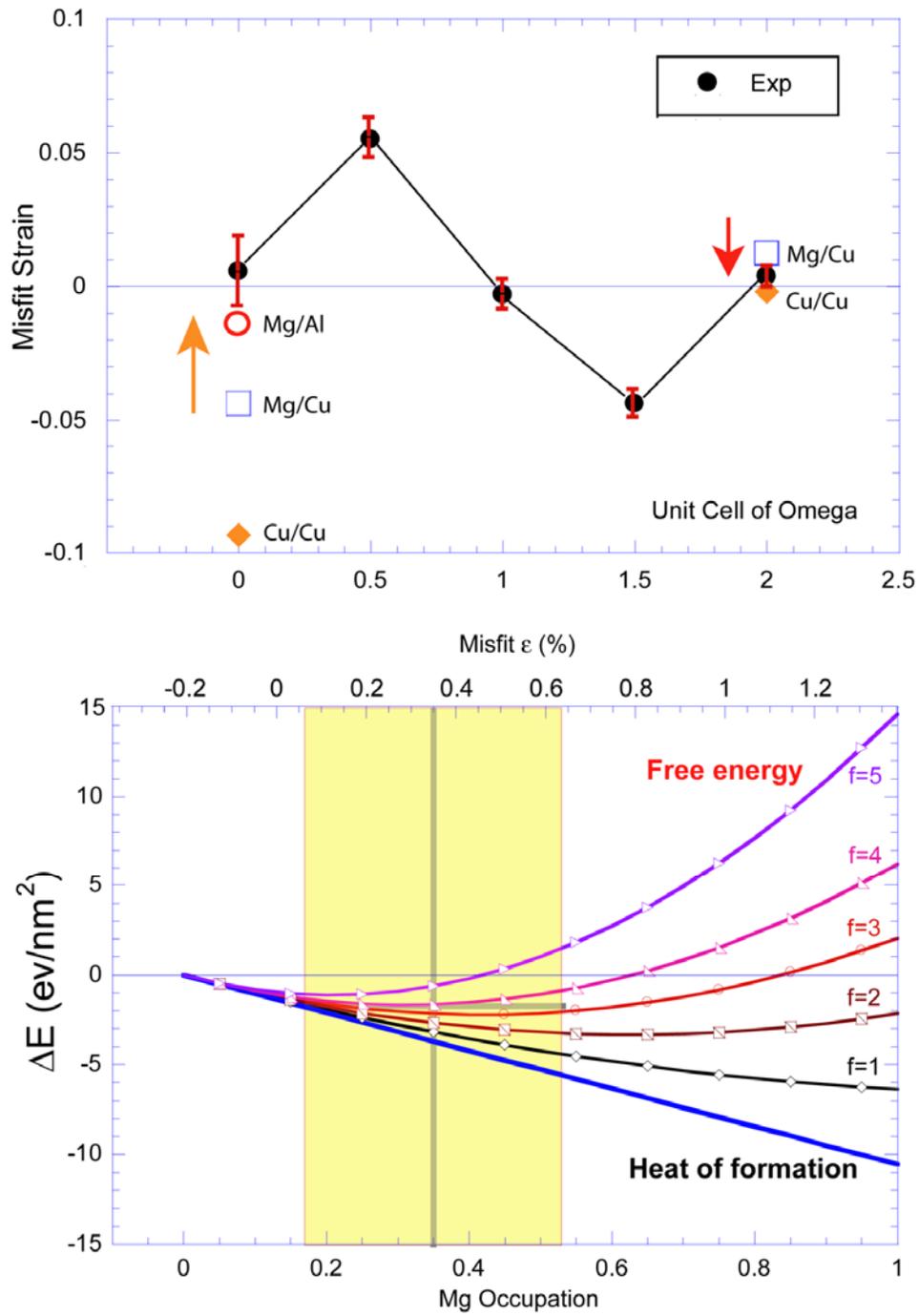

Figure 5